\documentclass{article}

     \usepackage[final,nonatbib]{neurips_2020}

\usepackage[utf8]{inputenc} % allow utf-8 input
\usepackage[T1]{fontenc}    % use 8-bit T1 fonts
\usepackage{microtype}      % microtypography

\usepackage{epsfig,graphicx,verbatim,parskip,tabularx,setspace,xspace}
\usepackage{booktabs}
\usepackage[font={small}]{caption}
\usepackage{multirow}
\usepackage{amssymb}

\title{Using ontology embeddings for structural inductive bias in gene expression data analysis}

\author{%
   Maja Tr\k{e}bacz    \quad\quad
   Zohreh Shams    \quad\quad
   Mateja Jamnik   \quad\quad
   Paul Scherer\\
   \textbf{Nikola Simidjievski\quad\quad
   Helena Andres Terre\quad\quad
   Pietro Li\`o} \\ 
   Department of Computer Science and Technology \\
   University of Cambridge, UK 
}

\begin{document}

\maketitle

\begin{abstract}
 Stratifying cancer patients based on their gene expression levels allows improving diagnosis, survival analysis and treatment planning. However, such data is extremely highly dimensional as it contains expression values for over 20000 genes per patient, and the number of samples in the datasets is low. To deal with such settings, we propose to incorporate prior biological knowledge about genes from ontologies into the machine learning system for the task of patient classification given their gene expression data. We use ontology embeddings that capture the semantic similarities between the genes to direct a Graph Convolutional Network, and therefore sparsify the network connections. We show
this approach provides an advantage for predicting  clinical targets from high-dimensional  low-sample data. 
\end{abstract}

\section{Introduction}

Ontologies are structured representations of semantic knowledge commonly used to represent biological concepts. For example, the Gene Ontology (GO)~\cite{ashburner2000gene}  unifies the representation of genes and their functions across all species. Ontologies prove useful for developing machine learning systems operating on biological entities~\cite{kulmanov2020machine, crawford2020incorporating, zitnik2019machine}, and they have an advantage of integrating data from
all the omics (genomes, transcriptomics, proteomics, metabolomics, etc.).
Recently developed ontology embeddings \cite{dl2vec, opa2vec, onto2vec, el} are structure-preserving maps from ontologies into vector spaces. The embeddings capture the semantic similarity between the genes. For example, genes \texttt{TP53} and \texttt{FOXA1} are close in the ontology space as they are transcription factors annotated to common molecular functions and biological processes, including terms related to cancer. We incorporate such prior knowledge into a machine learning system that classifies cancer subtypes from gene expressions.

The development of novel deep learning models on graphs \cite{belkinspectral, chebnet, gcn, mpnn} combined with biological motivations behind the network propagation model \cite{Cowen2017} has created an opportunity to leverage the prior biomedical knowledge without overly restricting the expressive power of the neural model. Rhee et al.~\cite{hybridrhee} proposed a combination of graph neural network (GNN) that uses a topology of protein-protein interaction (PPI) with a relation network~\cite{relationnetwork} to perform convolutions on the gene expressions and classify cancer patients into PAM50 subtype~\cite{parker2009supervised}. 
A similar work of Dutil et al.~\cite{Dutil2018TowardsGE} explores the usage of Graph Convolutional Network (GCN) directed by PPI networks~\cite{warde2010genemania} on the single gene inference tasks.  The follow-up works by Bertin et al.~\cite{Bertin2019AnalysisOG} and Hashir et al.~\cite{hashir2019graph} concluded that gene expression dependencies can be predicted almost as well
by using random networks as by using biological networks. One of the reasons for this is that the degree of the nodes varies significantly and the predictions tend to be worse for the genes with a low number of neighbours (or no neighbours at worst). Such nodes get less signal and thus tend to lead to lower performance. Crawford et al.~\cite{crawford2020graph} explored this observation and showed that when low-degree genes are removed, then the biological networks yield better performance than random graphs.

This work takes a different approach and is the first to use the ontology embeddings as a similarity measure between the genes in order to impose the inductive bias for the patient classification tasks from gene expressions.  We extend the gene expression convolution work of Dutil et al.~\cite{Dutil2018TowardsGE} by proposing novel graphs automatically generated from ontology embeddings, and a  node embedding method combining knowledge-based gene embeddings
with the expression values. Automatic generation of graphs expressing semantic similarity between genes allows overcoming the problems with nodes that have a small number of neighbours. Moreover, the ontology-based feature selection allows selecting a biologically relevant set of features.
% Another advantage of using ontologies is that they integrate
% all the omics (Genomes, transcriptomics,
% proteomics, metabolomcs etc) so they are good with data integration and
% they can also reach out clinical (clinical ontology) and exposure
% (environmental ontologies).

% On the other side, redutionist approach, the ontologies allow to follow
% “reasoning paths” to identify important biomarkers.

% Many CNN do now exploit all bioinformatics knowledge. With the
% incorporation of ontologies we use the "connectome" of all omics data.

\section{Methods}

We built the OntoGCN (ontology-directed Graph Convolutional Network) neural model where known similarities/relationships between the features (genes) direct processing in the network and help the model to avoid learning spurious correlations~\cite{Dutil2018TowardsGE}. OntoGCN enforces convolutions on the genes related by similarity and thus captures localised patterns of data, similarly as convolutional neural networks capture spatial relationships of pixels in the images~\cite{krizhevsky2012imagenet}. 

\begin{figure}[t!]
    \centering
    \includegraphics[width=\textwidth]{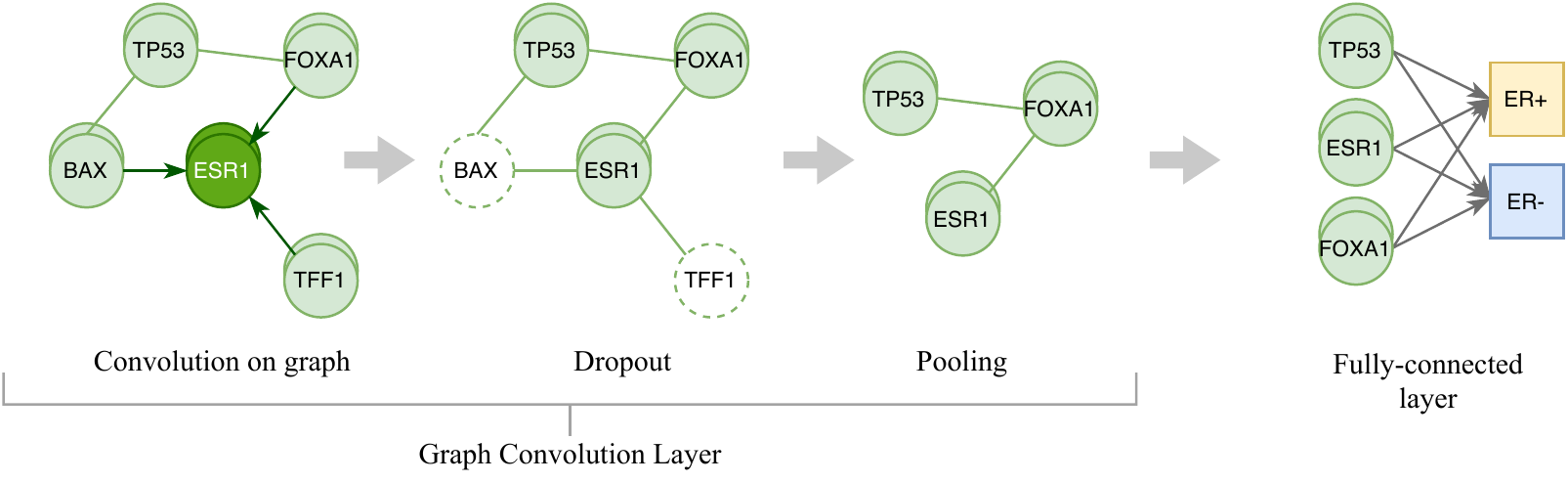}
    \caption{Overview of the OntoGCN  applied to gene expression data for patient classification.} 
    % (based on Dutil et al.~\cite{Dutil2018TowardsGE}). 
         \vskip -0.5cm
    \label{fig:graph_conv}
\end{figure}
Figure~\ref{fig:graph_conv} illustrates the workflow of our OntoGCN neural model. Each gene contributes a node to a  knowledge graph
where edges represent the similarity between the genes. 
We create edges that connect each gene to its K-nearest neighbours in the ontology embedding space according to the cosine distance.  We use DL2vec ontology embeddings that are a graph-based method  learning gene representations over three biomedical ontologies (GO \cite{ashburner2000gene}, UBERON \cite{mungall2012uberon}, and MP \cite{smith2009mammalian}).  The topology of the graph is the same for all the patient samples in the dataset, but each patient spans a new instance of the graph with different expression values at the nodes. At each graph convolution step, the neighbouring nodes are aggregated together based on their connectivity (Figure~\ref{fig:graph_conv} shows the update step for the ESR1 node, but similar message-passing happens for every node). After convolution, the gene nodes are dropped at random and also pooled using topology-based aggregation clustering.  Finally, a prediction is made from the remaining nodes via a fully connected layer.

A possible advantage of using a graph generated based on similarity, rather than a curated PPI network, is that one can freely control the sparsity or the number of neighbours for each of the nodes. This overcomes the problems regarding the nodes with few neighbours~\cite{hashir2019graph, Bertin2019AnalysisOG}. Moreover, we can generate a connected graph for any subset of genes present in the ontology embedding space, while taking a subset of genes for PPI network may result in a disconnected graph.

Dutil et al.~\cite{Dutil2018TowardsGE} learn embeddings for the genes during the training of the graph neural network. The input for the nodes in the model is these embeddings (initialised at random) scaled by their corresponding expression level for a particular sample. Such a method may potentially obfuscate the values of the input features (expressions).

The simplest way to avoid this is to use the expression values directly as the model input without scaling them via node embeddings. We propose another method of combining knowledge-based gene embeddings with the expression values (varying for each sample). First, the ontology embedding is concatenated with the expression and passed through a fully connected layer. Then we concatenate the expression value to the produced representation. Such an approach is potentially better than a simple scaling as it adjusts the node embeddings depending on the expression, and it also ensures the preservation of the expression value itself. 
Having knowledge-driven node embeddings helps convolution kernels to differentiate better between the genes during the graph convolutions, while at the same time it captures the similarity between the genes related by the biological knowledge.

\section{Experiments and results}

The experiments consider the tasks of classifying cancer patients from their genomic data collected by Molecular Taxonomy of Breast Cancer International Consortium (METABRIC) \cite{curtis2012genomic}. The classifications aim at three different subtypes of breast cancer: PAM50 (5-class molecular cancer subtypes) \cite{parker2009supervised}, ER (binary classification into immunohistochemistry subtypes) and iC10 (11 IntegrativeCluster subtypes) \cite{curtis2012genomic}. 

For the bulk of the implementation, we reused the codebase from Dutil et al.~\cite{Dutil2018TowardsGE}. We extended the implementation by custom graph topology generated from ontology embeddings and novel methods of generating knowledge-driven node embeddings. As the baselines, we use the Multi-Layer Perceptron (MLP) with dropout and Random Forest (RF).

\begin{table}[t!]
    \centering    
     \vskip -0.5cm
    \caption{Performance comparison of the methods on the PAM50, ER and iC10 patient classification tasks (accuracy$\pm$std), using all genes as input features. 
    Bold font marks the cases when the model obtained statistically significantly better results than the baselines.
    }
    \vskip 0.2cm

\resizebox{\textwidth}{!}{%
\begin{tabular}{rrcc|cc|cc}
\toprule
& & \multicolumn{2}{c}{ PAM50 }& \multicolumn{2}{c}{ ER } & \multicolumn{2}{c}{ IC10 }  \\
& train size & 100 &1500 & 100 &1500 & 100 & 1500 \\
\midrule

\multicolumn{2}{c}{OntoGCN w/o node embeddings }  & \textbf{72.3$\pm$2.9} & \textbf{81.2$\pm$0.8} & \textbf{91.4$\pm$1.0} & 93.7$\pm$1.0 & 48.2$\pm$3.1 & \textbf{74.3$\pm$2.3 }\\
% & random graph & 68.9$\pm$3.9 & 78.0$\pm$4.7 \\
% & genemania graph & 71.7$\pm$2.6 & 80.1$\pm$2.1 \\
% & stringdb graph & 71.2$\pm$2.5 & 80.7$\pm$1.2 \\
\multicolumn{2}{c}{OntoGCN w/ node embeddings} & \textbf{72.7$\pm$3.7} &\textbf{81.6$\pm$2.2} &90.8$\pm$0.5 & 93.8$\pm$1.2 &     50.4$\pm2$.3 & 73.7$\pm$3.3  \\
\midrule
\multicolumn{2}{c}{ Random Forest } & 69.3$\pm$1.1 & 78.4$\pm$1.0 & 88.7$\pm$1.2 & 93.1$\pm$1.2 & 66.7$\pm$0.9 & 71.3$\pm$2.1\\
\multicolumn{2}{c}{ Multi-Layer Perceptron } & 60.1$\pm$5.0 & 77.9$\pm$2.5 & 88.8$\pm$1.8 & 90.9$\pm$5.0 & 40.4$\pm$4.5 & 68.9$\pm$1.6  \\
\bottomrule
\end{tabular}
 }

     \vskip -0.5cm
    \label{tab:all_genes}
\end{table}

Table~\ref{tab:all_genes} presents the results in which all of the 24368 available genes are used as an input to the model with no feature selection. The evaluation explores various scenarios of data scarcity, as each experiment is performed using the training sample of 1500 or 100 patients.  

The results for PAM50 cancer subtype classification, presented in Table~\ref{tab:all_genes}, demonstrate a statistically significant improvement over the baseline models not using prior biological knowledge (with a p-value of 0.035 between the results of GCN and MLP for the training sample of 1500). 
The improvement achieved by GCN over baselines is especially visible in the case of the low data scenario of 100 patient samples. The MLP model is struggling to achieve high scores with the scarce data, due to the vast spurious connections. Directing the neural connections via ontology-based graph convolutions seems to overcome these problems and performs better. On the ER task, the GCN (with no node embeddings or the proposed ontology-based node embeddings) outperforms the baseline models in the low training data setting of 100 patients. On the task of iC10 classification, the GCN outperforms baselines when there is sufficient training data used. However, it struggles when trained on scarce data and applied to multi-label classification tasks such as iC10 (with 11 labels).

\begin{table}[b!]
 \vskip -0.5cm
    \centering
        \caption{Performance comparison of the graphs directing GCN structure  on the PAM50 classification task (accuracy $\pm$ std). The GCN used expression values as the nodes' input, without the node embedding mechanism.}
    \label{tab:topology_results}
         \vskip 0.2 cm
     \resizebox{0.56\textwidth}{!}{%
\begin{tabular}{lcc}
\toprule
&  \multicolumn{2}{c}{ PAM 50 } \\
& train size=100 & train size=1500 \\

\midrule
OntoGCN & \textbf{72.3$\pm$2.9} & \textbf{81.2$\pm$0.8}\\
Random graph & 68.9$\pm$3.9 & 78.0$\pm$4.7 \\
 GeneMANIA graph \cite{warde2010genemania} & 71.7$\pm$2.6 & 80.1$\pm$2.1 \\
 STRINGdb graph \cite{szklarczyk2019string} & 71.2$\pm$2.5 & 80.7$\pm$1.2 \\
 \midrule
 Random Forest  & 69.3$\pm$1.1 & 78.4$\pm$1.0\\
 Multi-Layer Perceptron  & 60.1$\pm$5.0 & 77.9$\pm$2.5   \\
\bottomrule
\end{tabular}
 }
%   \vskip 0.1cm

%         
\end{table}

In relation to the previous work on imposing bias from biological networks via graph convolutions, we compare in our experiments the graph generated using ontology embeddings with two graph datasets
containing a mixture of protein-protein interaction and gene
co-expression data:  GeneMANIA \cite{warde2010genemania} and STRINGdb \cite{szklarczyk2019string}. We also consider a baseline of a graph with randomly generated edges (with matching degree). Such a baseline allows determining if performance gains come from the model itself or the underlying prior biological knowledge.
% To relate the results with previous work by Dutil et al. \cite{Dutil2018TowardsGE} and Bertin et al. \cite{Bertin2019AnalysisOG}, we compare the automatically generated ontology graph versus topology of gene interaction networks (see Section \ref{graphs_used}). 
The results in Table~\ref{tab:topology_results} suggest that curated biological networks are a good source of prior knowledge. The slightly improved performance in case of the automatically generated ontology graph is probably caused by the ability to freely tune the sparsity of the graph, which regularises the sizes of kernels in convolution and overcomes the problems caused by sparsely connected genes in the network~\cite{Bertin2019AnalysisOG, crawford2020graph}.

We further combine the proposed GCN model with ontology-based feature selection strategy. The method combines prior knowledge of known cancer drivers with automatic selection based on semantic similarity of genes from the ontology embeddings. It begins by choosing a single gene that is known for its expression levels being highly correlated with the task. For example, for the task of predicting PAM50 subtype, such gene is ESR1, which encodes an estrogen receptor~\cite{bastien2012pam50, holst2007estrogen, oesterreich2013search}. Given at least one gene expression known as an important factor in the cancer-related classification task, one can select a set of (e.g., 1000) genes related with similar functions to the reference gene according to the cosine distance in the ontology embedding space.

\begin{figure}[t!]
    \centering
    \includegraphics[width=13cm]{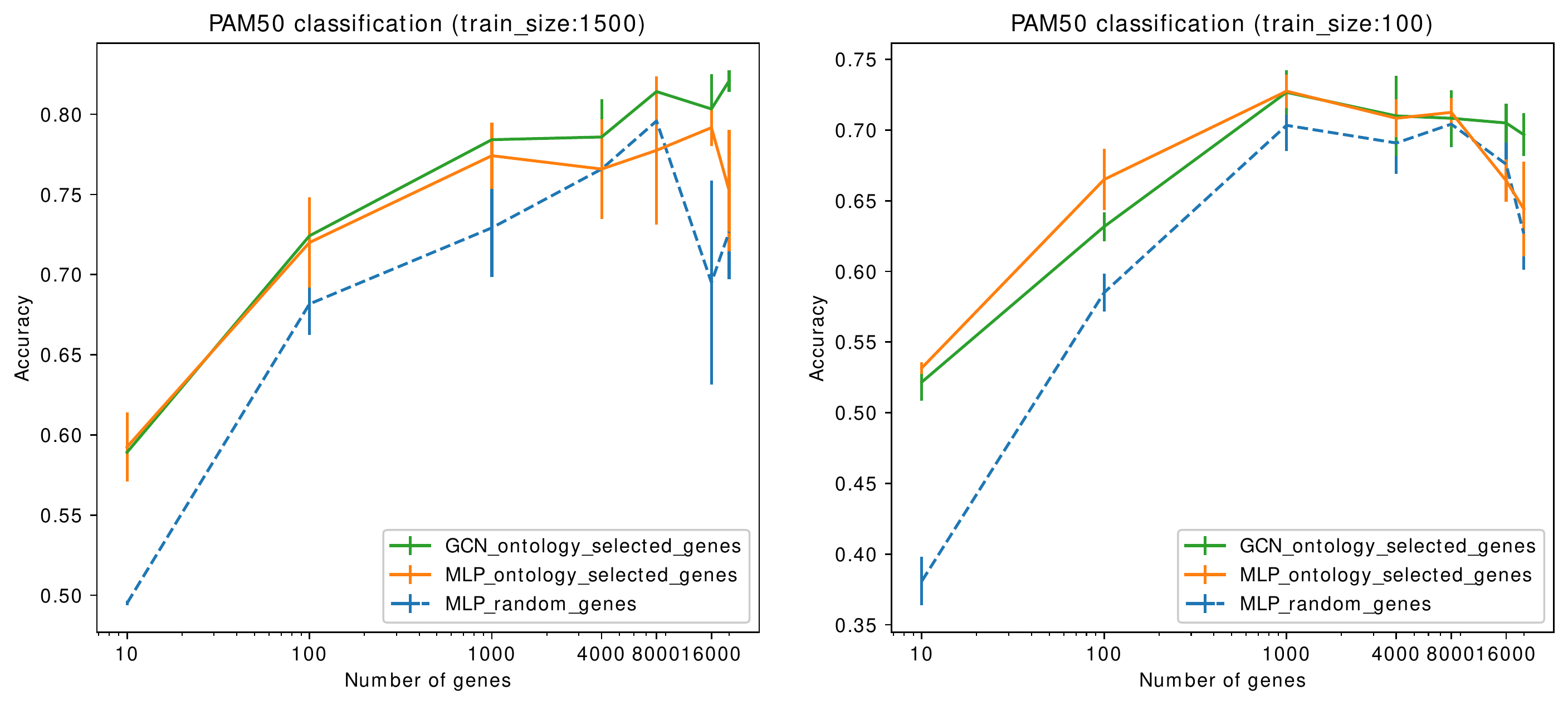}
    % \caption{Caption}
    % \label{fig:my_label}
% \end{figure}
% \begin{figure}[h!]
    \centering
    \caption{Accuracy obtained on the PAM50 breast cancer subtype classification tasks. The reported accuracy is an average over 5 trials (with different stratified split of data) and error bars represent standard error.  }
    \label{fig:gcn_with_fs_er_pam50}
\end{figure}
The results in Figure~\ref{fig:gcn_with_fs_er_pam50} for PAM50 classification show that the feature selection is important when using a smaller set of genes of cardinality below 4000. Given the genes chosen with the proposed feature selection strategy, both MLP and GCN outperform the baseline of MLP operating on a random set of genes. The differences are less significant on the larger gene-set covering most of the genes as the models can then discover signals distributed among a variety of genes.

The proposed GCN model with structural inductive bias based on ontology prior knowledge performs on par with MLP on the smaller selected set of genes. However, it can outperform the MLP in the situation when the whole high-dimensional input is considered. In such a situation the MLP (green) performance drops while GCN (yellow) continues to improve with a larger gene-set when given a large enough training sample. For PAM50 on scarce training samples, the performance of GCN and MLP reaches its peak around an input set of 1000 genes. After that MLP performance declines while GCN is still performing well.

\section{Conclusions}

We presented OntoGCN, a graph neural network model that exploits prior biological knowledge from ontologies to analyse
gene expression data in the cancer patient classification tasks. The model uses ontology embeddings \cite{dl2vec} to generate a graph directing the convolution operations in the network. 

The results suggest that imposing ontology-based knowledge as structural inductive bias in the model helps to mitigate the difficulties of high-dimensionality, by limiting the number of connections in the neural network architecture.  This can be useful for clinical tasks when little is known about the influence of different genes to a particular target bio-marker. In such scenarios, where all available data has to be put in use, the structural inductive that limits spurious connections proves beneficial.
However, if there is enough knowledge about the task to do feature selection, then  limiting the number of inputs and applying a simpler machine learning model may be a better way forward.

While in this paper we focus only on a number of clinical tasks that are related to known gene expressions, the proposed approach is general and applicable to other tasks.  Therefore, we intend to investigate the generality using other data from other clinical studies. Moreover, it would be particularly interesting to evaluate the method on clinical targets derived from heterogeneous data sources (such as image data as well as other types of genomic data).

\newpage

\bibliographystyle{unsrt} 

\bibliography{ref} 

\begin{thebibliography}{10}

\bibitem{ashburner2000gene}
Michael Ashburner, Catherine~A Ball, Judith~A Blake, David Botstein, Heather
  Butler, J~Michael Cherry, Allan~P Davis, Kara Dolinski, Selina~S Dwight,
  Janan~T Eppig, et~al.
\newblock Gene ontology: tool for the unification of biology.
\newblock {\em Nature genetics}, 25(1):25--29, 2000.

\bibitem{kulmanov2020machine}
Maxat Kulmanov, Fatima~Zohra Smaili, Xin Gao, and Robert Hoehndorf.
\newblock Machine learning with biomedical ontologies.
\newblock {\em bioRxiv}, 2020.

\bibitem{crawford2020incorporating}
Jake Crawford and Casey~S Greene.
\newblock Incorporating biological structure into machine learning models in
  biomedicine.
\newblock {\em Current Opinion in Biotechnology}, 63:126--134, 2020.

\bibitem{zitnik2019machine}
Marinka Zitnik, Francis Nguyen, Bo~Wang, Jure Leskovec, Anna Goldenberg, and
  Michael~M Hoffman.
\newblock Machine learning for integrating data in biology and medicine:
  Principles, practice, and opportunities.
\newblock {\em Information Fusion}, 50:71--91, 2019.

\bibitem{dl2vec}
Jun Chen, Azza~Th Althagafi, and Robert Hoehndorf.
\newblock Predicting candidate genes from phenotypes, functions, and anatomical
  site of expression.
\newblock {\em Bioinformatics}, 2020.

\bibitem{opa2vec}
Fatima~Zohra Smaili, Xin Gao, and Robert Hoehndorf.
\newblock Opa2vec: combining formal and informal content of biomedical
  ontologies to improve similarity-based prediction.
\newblock {\em Bioinformatics}, 35(12):2133--2140, 2019.

\bibitem{onto2vec}
Fatima~Zohra Smaili, Xin Gao, and Robert Hoehndorf.
\newblock Onto2vec: joint vector-based representation of biological entities
  and their ontology-based annotations.
\newblock {\em Bioinformatics}, 34(13):52--60, 2018.

\bibitem{el}
Maxat Kulmanov, Wang Liu-Wei, Yuan Yan, and Robert Hoehndorf.
\newblock El embeddings: geometric construction of models for the description
  logic el++.
\newblock pages 6103--6109, 2019.

\bibitem{belkinspectral}
Mikhail Belkin and Partha Niyogi.
\newblock Laplacian eigenmaps and spectral techniques for embedding and
  clustering.
\newblock In {\em Proceedings of the 14th International Conference on Neural
  Information Processing Systems: Natural and Synthetic}, NeurIPS'01, pages
  585--591. MIT Press, 2001.

\bibitem{chebnet}
Micha\"{e}l Defferrard, Xavier Bresson, and Pierre Vandergheynst.
\newblock Convolutional neural networks on graphs with fast localized spectral
  filtering.
\newblock In {\em Proceedings of the 30th International Conference on Neural
  Information Processing Systems}, NeurIPS'16, pages 3844--3852. Curran
  Associates Inc., 2016.

\bibitem{gcn}
Thomas~N. Kipf and Max Welling.
\newblock Semi-supervised classification with graph convolutional networks.
\newblock In {\em Proceedings of the 5th International Conference on Learning
  Representations}, ICLR'17, 2017.

\bibitem{mpnn}
Justin Gilmer, Samuel~S. Schoenholz, Patrick~F. Riley, Oriol Vinyals, and
  George~E. Dahl.
\newblock Neural message passing for quantum chemistry.
\newblock In {\em Proceedings of the 34th International Conference on Machine
  Learning - Volume 70}, ICML'17, pages 1263--1272, 2017.

\bibitem{Cowen2017}
Lenore Cowen, Trey Ideker, Benjamin~J. Raphael, and Roded Sharan.
\newblock Network propagation: a universal amplifier of genetic associations.
\newblock {\em Nature Reviews Genetics}, 18(9):551--562, Sep 2017.

\bibitem{hybridrhee}
Sungmin Rhee, Seokjun Seo, and Sun Kim.
\newblock Hybrid approach of relation network and localized graph convolutional
  filtering for breast cancer subtype classification.
\newblock In {\em Proceedings of the 27th International Joint Conference on
  Artificial Intelligence}, IJCAI’18, page 3527–3534. AAAI Press, 2018.

\bibitem{relationnetwork}
Adam Santoro, David Raposo, David~G Barrett, Mateusz Malinowski, Razvan
  Pascanu, Peter Battaglia, and Timothy Lillicrap.
\newblock A simple neural network module for relational reasoning.
\newblock In I.~Guyon, U.~V. Luxburg, S.~Bengio, H.~Wallach, R.~Fergus,
  S.~Vishwanathan, and R.~Garnett, editors, {\em Advances in Neural Information
  Processing Systems}, volume~30, pages 4967--4976, 2017.

\bibitem{parker2009supervised}
Joel~S Parker, Michael Mullins, Maggie~CU Cheang, Samuel Leung, David Voduc,
  Tammi Vickery, Sherri Davies, Christiane Fauron, Xiaping He, Zhiyuan Hu,
  et~al.
\newblock Supervised risk predictor of breast cancer based on intrinsic
  subtypes.
\newblock {\em Journal of clinical oncology}, 27(8):1160, 2009.

\bibitem{Dutil2018TowardsGE}
Francis Dutil, Joseph~Paul Cohen, Martin Weiss, Georgy Derevyanko, and Yoshua
  Bengio.
\newblock Towards gene expression convolutions using gene interaction graphs.
\newblock In {\em International Conference on Machine Learning (ICML) Workshop
  on Computational Biology (WCB)}, 2018.

\bibitem{warde2010genemania}
David Warde-Farley, Sylva~L Donaldson, Ovi Comes, Khalid Zuberi, Rashad
  Badrawi, Pauline Chao, Max Franz, Chris Grouios, Farzana Kazi,
  Christian~Tannus Lopes, et~al.
\newblock The genemania prediction server: biological network integration for
  gene prioritization and predicting gene function.
\newblock {\em Nucleic acids research}, 38(suppl\_2):W214--W220, 2010.

\bibitem{Bertin2019AnalysisOG}
Paul Bertin, Mohammad Hashir, Martin Wei{\ss}, Genevi{\`e}ve Boucher, Vincent
  Frappier, and Joseph~Paul Cohen.
\newblock Analysis of gene interaction graphs for biasing machine learning
  models.
\newblock {\em arXiv: Genomics}, abs/1905.02295, 2019.

\bibitem{hashir2019graph}
Mohammad Hashir, Paul Bertin, Martin Weiss, Vincent Frappier, Theodore Perkins,
  Genevi{\`e}ve Boucher, and Joseph~Paul Cohen.
\newblock Is graph biased feature selection of genes better than random?
\newblock {\em Machine Learning in Computational Biology (MLCB) meeting}, 2019.

\bibitem{crawford2020graph}
Jake Crawford and Casey~S Greene.
\newblock Graph biased feature selection of genes is better than random for
  many genes.
\newblock {\em BioRxiv}, 2020.

\bibitem{krizhevsky2012imagenet}
Alex Krizhevsky, Ilya Sutskever, and Geoffrey~E Hinton.
\newblock Imagenet classification with deep convolutional neural networks.
\newblock In {\em Advances in neural information processing systems}, pages
  1097--1105, 2012.

\bibitem{mungall2012uberon}
Christopher~J Mungall, Carlo Torniai, Georgios~V Gkoutos, Suzanna~E Lewis, and
  Melissa~A Haendel.
\newblock Uberon, an integrative multi-species anatomy ontology.
\newblock {\em Genome biology}, 13(1):R5, 2012.

\bibitem{smith2009mammalian}
Cynthia~L Smith and Janan~T Eppig.
\newblock The mammalian phenotype ontology: enabling robust annotation and
  comparative analysis.
\newblock {\em Wiley Interdisciplinary Reviews: Systems Biology and Medicine},
  1(3):390--399, 2009.

\bibitem{curtis2012genomic}
Christina Curtis, Sohrab~P Shah, Suet-Feung Chin, Gulisa Turashvili, Oscar~M
  Rueda, Mark~J Dunning, Doug Speed, Andy~G Lynch, Shamith Samarajiwa, Yinyin
  Yuan, et~al.
\newblock The genomic and transcriptomic architecture of 2,000 breast tumours
  reveals novel subgroups.
\newblock {\em Nature}, 486(7403):346--352, 2012.

\bibitem{szklarczyk2019string}
Damian Szklarczyk, Annika~L Gable, David Lyon, Alexander Junge, Stefan Wyder,
  Jaime Huerta-Cepas, Milan Simonovic, Nadezhda~T Doncheva, John~H Morris, Peer
  Bork, et~al.
\newblock String v11: protein--protein association networks with increased
  coverage, supporting functional discovery in genome-wide experimental
  datasets.
\newblock {\em Nucleic acids research}, 47(D1):D607--D613, 2019.

\bibitem{bastien2012pam50}
Roy~RL Bastien, {\'A}lvaro Rodr{\'\i}guez-Lescure, Mark~TW Ebbert, Aleix Prat,
  Blanca Mun{\'a}rriz, Leslie Rowe, Patricia Miller, Manuel Ruiz-Borrego,
  Daniel Anderson, Bradley Lyons, et~al.
\newblock Pam50 breast cancer subtyping by rt-qpcr and concordance with
  standard clinical molecular markers.
\newblock {\em BMC medical genomics}, 5(1):44, 2012.

\bibitem{holst2007estrogen}
Frederik Holst, Phillip~R Stahl, Christian Ruiz, Olaf Hellwinkel, Zeenath
  Jehan, Marc Wendland, Annette Lebeau, Luigi Terracciano, Khawla Al-Kuraya,
  Fritz J{\"a}nicke, et~al.
\newblock Estrogen receptor alpha (esr1) gene amplification is frequent in
  breast cancer.
\newblock {\em Nature genetics}, 39(5):655--660, 2007.

\bibitem{oesterreich2013search}
Steffi Oesterreich and Nancy~E Davidson.
\newblock The search for esr1 mutations in breast cancer.
\newblock {\em Nature genetics}, 45(12):1415--1416, 2013.

\end{thebibliography}

\end{document}